\documentclass[a4paper,12pt]{article}
\usepackage[utf8]{inputenc}
\usepackage{authblk}
\usepackage{graphicx}
\usepackage{xcolor}
\usepackage{subfigure}
\usepackage{amsmath}
\usepackage{amsbsy}

\usepackage[
maxnames=3, 
sorting=none, 
backend=bibtex,
style=numeric,
]{biblatex}
\newcommand{\der}{\mathrm{d}}
\newcommand{\dert}{\mathrm{d}t}
\newcommand{\ddert}{\frac{\mathrm{d}}{\mathrm{d}t}}
\newcommand{\dderr}{\frac{\mathrm{d}}{\mathrm{d}r}}

\def\bf#1{\mathrm{\mathbf{#1}}}
\def\bfg#1{\pmb{#1}}

\title{Implementation and validation of the relativistic transient absorption theory within the dipole approximation}

\author[$\dagger$]{Felipe Zapata}
\author[$\dagger$,$\ddag$]{Jimmy Vinbladh}
\author[$\ddag$]{Eva Lindroth}
\author[$\dagger$]{Jan Marcus Dahlstr\"om}

\affil[$\dagger$]{\small Department of Physics, Lund University, 22100 Lund, Sweden.}
\affil[$\ddag$]{\small  Department of Physics, Stockholm University, 10691 Stockholm, Sweden.}

\date{}

\begin{document}

\maketitle

\begin{abstract}
 A relativistic transient absorption theory is derived, implemented and validated within the dipole approximation based on the time-dependent Dirac equation. 
 In the non-relativistic limit, it is found that the absorption agrees  with the well established non-relativistic theory based on the time-dependent Schrödringer equation. 
 Time-dependent simulations have been performed using the Dirac equation and the Schrödinger equation for the Hydrogen atom in two different attosecond transient absorption scenarios. These simulations validate the present relativistic theory. The presented work can be seen as a first step in the development of a more general relativistic attosecond transient absorption spectroscopy method for studying heavy atoms, but it also suggests the possibility of studying relativistic effects, such as \emph{Zitterbewegung}, in the time domain.
\end{abstract}

\section{Introduction}
Femtochemistry has opened the possibility of studying molecular motion thanks to the development of ultra-short laser pulses \cite{Zewail00}. Attophysics is a natural continuation of femtochemistry that aims to unravel the electron motion in atoms and molecules by using attosecond laser pulses of coherent extreme ultraviolet (XUV) radiation \cite{Misha09}. For that reason, many experimental ideas have been transfer from the femtosecond to the attosecond domain. An example of this is attosecond transient absorption spectroscopy (ATAS),  which is used to study electron coherence and motion in atoms \cite{Wirth11,Goulielmakis10,Sabbar17}. Typically, an intense laser field is used to prepare an electron wave packet in a specific state of the target atom, while an XUV pulse is used to probe the dynamics. In addition, different scenarios have been proposed to investigate effects such as the Aulter-Townes splitting and the control of Fano line shapes \cite{Wang10,Ott13,Ott14}. Recent ATAS experiments have resolved light-induced structures in the photoabsorption spectrum, below and above the ionization threshold, in the Helium atom \cite{Birk20}. A general overview of the experimental work is given in Ref. \cite{Beck15}. The time-dependent Schr\"odinger equation (TDSE) has been used to study autoionization states in noble gases \cite{Chu13,Petersson17,Chew18}. A review on derivation of strong field ATAS, together with related phenomena, can be found in Ref. \cite{Wu16}.  

All studies so far are based on \emph{non-relativistic} ATAS theory \cite{Wu16}, but with spin-orbit corrections added in some cases. The importance of spin-orbit coupling was demonstrated already by the first ATAS experiment, which targeted Krypton   \cite{Goulielmakis10}. Calculations including spin-orbit effects have been done by Pabst \emph{et al}  \cite{Pabst12} as well as by  Baggesen \emph{et al} \cite{Baggesen12}. Pabst \emph{et al} treated a strong near-infrared pump-pulse, and an overlapping XUV probe pulse, using  the time-dependent configuration interaction singles (TDCIS) approach with an \emph{ad hoc} Pauli-type spin-orbit correction on to the occupied states. Baggesen \emph{et al}, on the other hand, considered two attosecond pulses both weak enough that the light-matter interaction could be treated by the lowest-order perturbation theory. A fully relativistic many-body approach was used to describe the bound states and hole transitions, while a non-relativistic discrete continuum was used to describe the ejected electron. Transitions between continuum states were not considered.

Regarding the lack of a \emph{relativistic} transient absorption theory, our aim is to developed a method that can handle strong fields, far beyond the perturbative regime. We have thus derived a general relativistic transient absorption theory which is based on the time-dependent Dirac equation (TDDE). General speaking, it could be used to study heavy atomic elements and relativistic effects, such as the \emph{Zitterbewegung} effect \cite{Schrodinger30}. In the present implementation, the description of the light-matter interaction is restricted to the dipole approximation, which is sufficient for strong, albeit not extreme fields. Positron-electron pair creation is not investigated in this work.

The outline of the present article is as follows. In Section \ref{theory}, theory is presented. The instantaneous power absorbed by the atom is derived in both length and velocity gauges. The relativistic theory is an extension of the non-relativistic one. As a consequence, the \emph{non-relativistic limit} of the theory based on the TDDE should give the same results as the well established non-relativistic ATAS theory, enabling a reference point for comparison. Thus, TDDE and TDSE calculations have been carried out on the Hydrogen atom due to its \emph{low relativistic nature}.  In Section \ref{numerical_method}, the procedure to solve numerically the TDDE is given and, in Section \ref{validation_results}, results are presented for two different ATAS scenarios. The first is based on a simple one-photon ionization process, while the second is a multi-photon ionization process generated by a non-linear pump and a linear probe laser pulse. Finally, in Section \ref{conclusion}, the conclusion is given. 

\section{Theory\label{theory}}
In ATAS, the target quantity is the exchange of energy between the atomic system and the laser field, $\Delta\mathcal{E}$. Thanks to energy conservation, one may infer that the absorbed energy from the laser field can be obtained by calculating the energy gained by the excited atom \cite{Wu16}. Formally, this latter energy can be expressed as
\begin{equation}
\label{energy}
\Delta \mathcal{E} = \int_{-\infty}^{+\infty}{\Delta\dot{\mathcal{E}}}(t)\dert,
\end{equation}
where the instantaneous power delivered to the atom is defined as
\begin{equation}
\label{power}
\Delta\dot{\mathcal{E}}(t)=\ddert\langle\psi(t)|H(t)|\psi(t)\rangle=\langle\psi(t)|\dot{H}(t)|\psi(t)\rangle,
\end{equation}                          
where $\psi(t)$ is the time-dependent wave function and $H(t)$ is the atomic Hamiltonian including interaction with external field. According to Ref.~\cite{dahlstrom_attosecond_2017}, the instantaneous power defined by Eq.~(\ref{power}) is an elusive quantity as it is not gauge invariant. This fact can be understood as a result of the time-dependent redefinition of the absolute energy scale, discussed in many textbooks on quantum mechanics ~\cite{Sakurai67}.
In order to remove the gauge ambiguity, we introduce a relative instantaneous power, $\Delta \dot {\bar{\mathcal{E}}}(t)$, defined as the difference between the atomic instantaneous power and the instantaneous power of a free electron, $\Delta\dot{\mathcal{E}}_0(t)$,
\begin{equation}
\label{gaugeinvariantpower}
    \Delta \dot {\bar{\mathcal{E}}}(t) = 
    \Delta \dot {\mathcal{E}}(t)  - \Delta \dot {\mathcal{E}}_0(t),
\end{equation} 
where both systems are subjected to the same time-dependent laser field described in the same gauge. In this way, the free electron serves as a physical reference system for the energy and power of the atomic system. As we will see, this subtraction will correct the time-dependent power to the atom without affecting the total energy gain. This is because $\Delta{\mathcal{E}}_0=0$ for a free electron interacting with a laser field, which is implemented by choosing a vector potential that vanishes at minus and plus infinity in time. 

In this section, the derivation of the relativistic instantaneous power is presented. As the relativistic theory can be seen as an extension of the non-relativistic theory, we shall first revisit the derivation of the non-relativistic instantaneous power. 

\subsection{Non-relativistic instantaneous power delivered to the atom}

The time-dependent Schrödinger equation (TDSE) encodes the non-relativistic electron dynamics and is written as follows
\begin{equation}
i\hbar\frac{\partial}{\partial t}\psi(t)=H^{S\Theta}(t)\psi(t). 
\end{equation}
In presence of a laser field, the time-dependent Hamiltonian $H^{S\Theta}(t)$ can be expressed, within the dipole approximation, either in the length $(SL)$ or in the velocity $(SV)$ gauge. Specifically, for a linear polarized laser field along the $z$-axis, one can write \cite{Sakurai67} 
\begin{eqnarray}
\label{SL}
H^{SL}(t)&=&H^S_0 + e z  E(t),\\
\label{SV}
H^{SV}(t)&=&H^S_0 + \frac{e}{m_e}p_z A(t) +\frac{e^2}{2m_e}A^2(t),
\end{eqnarray}
where $z$ is the electron position and $p_z$ represents the $z$-component of the momentum vector. The relation between the vector potential $A(t)$ and the electric field $E(t)$ is given by $E(t)=-\dot{A}(t)$. Finally, the field-free Hamiltonian is given by
\begin{equation}
    H^S_0=\frac{\bf{p}^2}{2m_e}-\frac{e^2Z}{4\pi\epsilon_0 r},
\end{equation}
where $\bf{p}=-i\hbar\boldsymbol{\nabla}$ is the momentum operator and the electron-nucleus interaction is described by the Coulomb potential.

\subsubsection{Length form} 
In order to compute the instantaneous power delivered to the atom in its length form,  Eq.(\ref{SL}) is used. Then, one can write 
\begin{eqnarray}
\label{power_SL}
\nonumber
\Delta\dot{\mathcal{E}}_\mathrm{TDSE}^{(L)}(t)&=&\langle\psi(t)|\dot{H}^{SL}(t)|\psi(t)\rangle\\
\nonumber
&=&e\langle\psi(t)|z|\psi(t)\rangle\;\dot{E}(t)\\
&=&e z(t)\dot{E}(t),
\end{eqnarray}             
where $z(t)$ is the expectation value of the electron position. Therefore, the integrated power (total gain energy) is expressed as 
\begin{equation}
\Delta \mathcal{E}^{(L)}_\mathrm{TDSE}=e\int_{-\infty}^{+\infty}z(t)\dot{E}(t)\dert.
\end{equation}

Following Wu \emph{et al} in Ref. \cite{Wu16}, the total gain energy can be rewritten in the energy domain as follows 
\begin{equation}
\Delta\mathcal{E}^{(L)}_\mathrm{TDSE}= - 2e \int_{0}^{+\infty}\omega \;\mathrm{Im}\{\tilde{z}(\omega)\tilde{E}^*(\omega)\} \der\omega,
\end{equation}
where $\tilde{z}(\omega)=\tilde{z}^*(-\omega)$ and $\tilde{E}(\omega)=\tilde{E}^*(-\omega)$ are the Fourier transforms of the real $z(t)$ and $E(t)$ functions, respectively. The energy-resolved power in the length gauge is identified as  
\begin{equation}
\label{ins_sl}
P_\mathrm{TDSE}^{(L)}(\omega)=-2e\omega\;\mathrm{Im}\{\tilde{z}(\omega)\tilde{E}^*(\omega)\},
\end{equation}
where the energy argument is positive, $\omega\ge0$.
Finally, the reference instantaneous power, defined for a free electron in length gauge is given by
\begin{equation}
\label{rpowerL}
\Delta \dot{\mathcal{E}}^{(L)}_0(t)=ez_0(t)\dot E(t),    
\end{equation}
where the free electron position $z_0(t)$ is computed as  
\begin{equation}
\label{freeposition}
z_0(t)=-\frac{e}{m}\int_{-\infty}^{t}dt' \int_{-\infty}^{t'} E(t'') dt'' =\frac{e}{m}\int_{-\infty}^{t} A(t') dt'.  
\end{equation}

\subsubsection{Velocity form}
The velocity form is obtained by making use of Eq.(\ref{SV}). Then, the instantaneous power is expressed as  
\begin{eqnarray}
\label{power_SV}
\nonumber
\Delta\dot{\mathcal{E}}_\mathrm{TDSE}^{(V)}(t)&=&\langle\psi(t)| \dot{H}^{SV}(t)|\psi(t)\rangle\\
 \nonumber
 &=&\frac{e}{m}\langle\psi(t)|p_z|\psi(t)\rangle\dot{A}(t)+\frac{e^2}{m}A(t)\dot{A}(t)\\
 &=&-\frac{e}{m}[ p_z(t)+eA(t)]E(t)
\end{eqnarray}      
where $p_z(t)$ is the expectation value of the $z$-component of the canonical momentum. We have to mention that the expectation value $p_z(t)$ is not gauge invariant. However, the expectation value of the $z$-component of the velocity operator, given by 
\begin{equation}
v_z(t)=\frac{e}{m}[p_z(t)+eA(t)],
\end{equation}
is gauge invariant, \emph{see} for example Ref. \cite{Sakurai67}. Subsequently, the integrated power can be computed as follows
\begin{equation}
\Delta \mathcal{E}^{(V)}_\mathrm{TDSE}= -\frac{e}{m}\int_{-\infty}^{+\infty}[p_z(t)+eA(t)]E(t)\dert.
\end{equation}
Following Ref. \cite{Marcus17}, the integrated power, in the energy domain, is given by
\begin{equation}
\label{energy_SV}
\Delta \mathcal{E}^{(V)}_\mathrm{TDSE}= - 2\frac{e}{m}\int_{0}^{+\infty}\omega \;\mathrm{Im}\{[\tilde{p}_z(\omega)+e\tilde{A}(\omega)]\tilde{A}^*(\omega)\} \der\omega,
\end{equation}
where $\tilde{p}_z(\omega)=\tilde{p}_z^*(-\omega)$ and $\tilde{A}(\omega)=\tilde{A}^*(-\omega)$ are the Fourier transforms of the real functions $p_z(t)$ and $A(t)$. This implies that the energy-resolved power is given by
\begin{equation}
\label{ins_sv}
P_\mathrm{TDSE}^{(V)}(\omega)=-2\frac{e}{m}\omega\;\mathrm{Im}\{\tilde{p}_z(\omega)\tilde{A}^*(\omega)\},
\end{equation}
where we used the fact that the term $\tilde{A}(\omega)\tilde{A}^*(\omega)=|\tilde{A}(\omega)|^2$ in Eq.~(\ref{energy_SV}) is real. 

Finally, the instantaneous  power delivered to a free electron, in velocity gauge, is given by
\begin{equation}
\label{rpowerV}
\Delta \dot{\mathcal{E}}^{(V)}_0(t)=-ev_0(t) E(t)
\end{equation}
where the free electron velocity in the laser field is $v_0(t)= eA(t)/m$.  

\subsection{Relativistic instantaneous power delivered to the atom\label{rel_ATAS}}
The relativistic dynamics of an electron embedded in the field of an atomic nucleus is described by the time-dependent Dirac equation (TDDE) which is defined as follows,
\begin{equation}
\label{TDDE}
i\hbar\frac{\partial}{\partial t}\psi(t)=H^{D\Theta}(t)\psi(t).
\end{equation}
 In the  presence of a polarized laser field and within the dipole approximation, the time-dependent Dirac Hamiltonian $H^{D\Theta}(t)$  is expressed in the length gauge ($DL$) as
\begin{equation}
\label{DL}
    H^{DL}(t)=H_0^D+e z E(t), 
\end{equation}
while the interaction with the laser field in the velocity gauge ($DV$) is given by 
\begin{equation}
\label{DV}
H^{DV}(t)=H_0^D+ec\;\alpha_z A(t),
\end{equation}
 \emph{see} for example Ref.~\cite{Vanne12,Jakob09,Kjellsson17,Kjellsson17b}.
Finally, the time-independent Dirac Hamiltonian for an electron in the potential generated by an infinite-mass, point-charge, nucleus  of charge $Z$ is 
\begin{equation}     
\label{FFD}
H_0^D= ec\;\bfg{\alpha} \cdot \bf{p} + m_e c^2\beta-\frac{e^2Z}{4\pi\epsilon_0 r},
\end{equation}
where $m_e c^2$ is the rest mass energy of the electron. The Dirac matrices $\boldsymbol{\alpha}=(\alpha_x,\alpha_y,\alpha_z)$ and $\beta$  are expressed as 
\begin{equation}
\nonumber
    \alpha_q=
    \left(\begin{array}{cc}
        0 & \sigma_q  \\
        \sigma_q & 0 
    \end{array}\right)\;\;\mathrm{and}\;\;\beta=
    \left(\begin{array}{cc}
        I & 0  \\
        0 & -I
    \end{array}\right),
\end{equation}
with
\begin{equation}
\nonumber
    \sigma_x= 
    \left(\begin{array}{cc}
        0 & 1  \\
        1 & 0 
    \end{array}\right),\;\;\sigma_y= 
    \left(\begin{array}{cc}
        0 & -i  \\
        i & 0 
    \end{array}\right),\;\;\mathrm{and}\;\;
    \sigma_z= 
    \left(\begin{array}{cc}
        1 & 0  \\
        0 & -1 
    \end{array}\right),
\end{equation}
where the set of $\sigma_q$ is given by the Pauli matrices and $I$ is a $2\times2$ unity matrix  \cite{Sakurai67,Grant06}. Since the Dirac matrices $\boldsymbol{\alpha}$ and $\beta$ are $4\times4$ matrices, the wave function in Eq.(\ref{TDDE}) is defined as a four component function.

\subsubsection{Length form}
The relativistic instantaneous power delivered to the atom can be computed in the length form using Eq.(\ref{DL}) as follows
\begin{eqnarray}
\label{power_DL}
\nonumber
\Delta\dot{\mathcal{E}}_\mathrm{TDDE}^{(L)}(t)&=&\langle\psi(t)|\dot{H}^{DL}(t)|\psi(t)\rangle\\
\nonumber
&=&e\langle\psi(t)|z|\psi(t)\rangle\;\dot{E}(t)\\
&=&ez(t)\dot{E}(t),
\end{eqnarray}    
where now the expectation value $z(t)$ is computed with the relativistic four component wave function. Thus, the length form of the integrated power is given by 
\begin{equation}
\Delta \mathcal{E}^{(L)}_\mathrm{TDDE}=e\int_{-\infty}^{+\infty}z(t)\dot{E}(t)\dert,
\end{equation}
which in the energy representation is
\begin{equation}
\label{EF_DL}
\Delta \mathcal{E}^{(L)}_\mathrm{TDDE}= - 2 e\int_{0}^{+\infty}\omega \;\mathrm{Im}\{\tilde{z}(\omega)\tilde{E}^*(\omega)\} \der\omega.
\end{equation}

As in the non-relativistic case, Eq.(\ref{EF_DL}) implies that the energy-resolved power is defined as 
\begin{equation}
\label{ins_rl}
P_\mathrm{TDDE}^{(L)}(\omega)=-2e\omega\;\mathrm{Im}\{\tilde{z}(\omega)\tilde{E}^*(\omega)\}.
\end{equation}
As we can see, the expression of the relativistic instantaneous power in the length form is similar to the corresponding expression in the non-relativistic theory. The important difference concerns the calculation of the time-dependent expectation value $z(t)$. 

\subsubsection{Velocity form}
The velocity form of the relativistic power is obtained using Eq.[\ref{DV}]
\begin{eqnarray}
\label{power_DV}
\nonumber
\Delta\dot{\mathcal{E}}_\mathrm{TDDE}^{(V)}(t)&=&\langle\psi(t)|\dot{H}^{DV}(t)|\psi(t)\rangle\\
\nonumber
&=&ec\;\langle\psi(t)|\alpha_z|\psi(t)\rangle\;\dot{A}(t)\\
&=&-ec\;\alpha_z(t)E(t),
\end{eqnarray}      
where $\alpha_z(t)$ is the expectation value of the $z$-component of the  $\bfg{\alpha}$ vector. Eq.(\ref{power_DV}) is the relativistic version of Eq.(\ref{power_SV}). In the relativistic theory, the $z$-component of the velocity operator is defined by $v_{z}=i\hbar[H^{DV},z]=ec\alpha_z$ \cite{Sakurai67}, and its expectation value is given by
\begin{equation}
    v_z(t)=ec\alpha_z(t).
\end{equation}
As in the non-relativistic theory, a corresponding classical expression for the power (velocity $\times$ force) is recovered. For more details on the interpretation of the one-particle relativistic operators, \emph{see} Ref. \cite{Greiner87}. 

Consequently, the integrated power is given by the following expression
\begin{equation}
\Delta \mathcal{E}^{(V)}_\mathrm{TDDE}= -ec\int_{-\infty}^{+\infty}\alpha_z(t)E(t)\dert,
\end{equation}
and its associated form in the energy domain is given by
\begin{equation}
\label{rpower_energy}
\Delta \mathcal{E}^{(V)}_\mathrm{TDDE}= - 2ec \int_{0}^{+\infty}\omega \;\mathrm{Im}\{\tilde{\alpha}_z(\omega)\tilde{A}^*(\omega)\} \der\omega,
\end{equation}
where $\tilde{\alpha}_z(\omega)=\tilde{\alpha}_z^*(-\omega)$  is the Fourier transform of the real function $\alpha_z(t)$.

Finally, the energy-resolved power transfer to the atom can be identified from Eq.(\ref{rpower_energy}) as 
\begin{equation}
\label{ins_rv}
P_\mathrm{TDDE}^{(V)}(\omega)=-2e\omega c\;\mathrm{Im}\{\tilde{\alpha}_z(\omega)\tilde{A}^*(\omega)\}.
\end{equation}

At this point, some comments can be done on the derivation of Eq.(\ref{ins_rl}) and Eq.(\ref{ins_rv}). We have established a formal relativistic transient absorption theory as an extension of the non-relativistic theory. Thus, at its non-relativistic limit, similar results should be obtain by Eq.~(\ref{ins_sl}) and Eq.~(\ref{ins_sv}), and by Eq.~(\ref{ins_rl}) and Eq.~(\ref{ins_rv}). In this work, the non-relativistic limit of the theory has been explored numerically in Section (\ref{validation_results}). In the next section, the numerical method used to solve the TDDE is presented. Atomic units are used unless otherwise stated, $e=\hbar=m_e=4\pi\epsilon_0=1$.

\section{Numerical solution of the time-dependent Dirac equation\label{numerical_method}}
The TDDE is solved numerically by using the spectral method for the propagation \cite{Vanne12,Kjellsson17,Kjellsson17b}. With this approach, the time-dependent wave function is expanded onto the field-free Dirac states $\{\Phi_K({\bf r})\}$ as follows,
\begin{equation}
\label{expansion}
    \psi(t)=\sum_K C_K(t)\Phi_K({\bf r}),
\end{equation}
where the set $\{C_K(t)\}$ contains the time-dependent coefficients. Hence, within the interaction picture, the TDDE can be propagated using the second-order finite-differencing scheme that has been used previously to propagate the TDSE \cite{Leforestier91}. In consequence,  time-dependent coefficients are computed at each time-step $Delta t$ as 
\begin{equation}
C_K(t+\Delta t)=e^{-2i\omega_K\Delta t}C_K(t-\Delta t)-2i\Delta t e^{i\omega_K \Delta t}F_K^{D\Theta}(t),
\end{equation}
where the ground state energy has been defined to be zero, $\omega_K=\varepsilon_K-\varepsilon_0$, following the implementation made by Greenman \emph{et al.} in Ref. \cite{Greenman10}.  The function $F_K^{D\Theta}(t)$ can be defined in the length and in the velocity gauge as follows
\begin{eqnarray}
\label{FKDL}
F_K^{DL}(t)&=&\sum_{K'}C_{K'}(t)M_{K',K}^{DL}(t), \\
\label{FKDV}
F_K^{DV}(t)&=&\sum_{K'}C_{K'}(t)M_{K',K}^{DV}(t),
\end{eqnarray}
where $M_{K',K}^{DL}(t)$ and $M_{K',K}^{DV}(t)$ are the time-dependent transition matrix elements,  $\{\varepsilon_K\}$  are the Dirac state energies. The norm is preserved and stability is obtained for $\Delta t < \hbar / |\varepsilon_\mathrm{max}|$, being $\varepsilon_\mathrm{max}$ the largest eigenvalue the Hamiltonian operator \cite{Leforestier91}. Finally, the time-dependent expectation value of the electron position and velocity can be computed as 
\begin{eqnarray}
\label{ZMKDL}
    z(t)=\sum_{K'}\sum_{K}C_{K'}^*(t)C_K(t)\tilde{M}_{K',K}^{DL}(t),\\
\label{VMKDV}
    v_z(t)=\sum_{K'}\sum_{K}C_{K'}^*(t)C_K(t)\tilde{M}_{K',K}^{DV}(t),
\end{eqnarray}
where the transition matrix elements $\tilde{M}_{K',K}^{DL}(t)$ and $\tilde{M}_{K',K}^{DV}(t)$ are, in principle, the same as the given in Eq.~(\ref{FKDL}) and in Eq.~(\ref{FKDV}), respectively. However, as we will show in the next section, if a complex absorbing method is used when solving the radial Dirac equations, matrix elements in Eq.~(\ref{ZMKDL}) and Eq.~(\ref{VMKDV}) will need to be evaluated in a different manner with respect to ${M}_{K',K}^{DL}(t)$ and ${M}_{K',K}^{DV}(t)$. 

\subsection{Transition matrix elements with CAP}
In order to compute the transition matrix elements appearing in Eq.~(\ref{FKDL}) and in Eq.~(\ref{FKDV}) required for propagation, and in Eq.~(\ref{ZMKDL}) and in Eq.~(\ref{VMKDV}) necessary for the calculation of the expectation values, one needs first to obtain the Dirac states $\{\Phi_K({\bf r})\}$. For a spherically symmetric $H_0^D$, the states can be written as four component spinors of the following form
\begin{equation}
\Phi_K(\bf{r})\equiv\Phi_{n,j,m,\kappa}(\bf{r})=\frac{1}{r}\binom{P_{n,\kappa}(r)X_{\kappa,j,m}(\Omega)}{iQ_{n,\kappa}(r)X_{-\kappa,j,m}(\Omega)},
\end{equation}
where  $\kappa=\ell$ for $j=\ell-1/2$ and $\kappa=-(\ell+1)$ for $j=\ell+1/2$. The spin-angular part $X_{\kappa,j,m}(\Omega)$ are computed analytically as a linear combination of products of spherical harmonics $Y_{\ell_\kappa}^{m_{\ell_\kappa}}(\Omega)$ and spin eigenstates $\chi_{m_s}$ as follows 
\begin{equation}
    X_{\kappa,j,m}(\Omega)=\sum_{m_\ell,m_s}\langle \ell_\kappa,m_s;s=1/2,m_s|j,m\rangle Y_{\ell_\kappa}^{m_{\ell_\kappa}}(\Omega)\chi_{m_s},
\end{equation}
where $\langle \ell_\kappa,m_s;s=1/2,m_s|j,m\rangle$ are the Clebsch-Gordan coefficients and $\Omega$ stands for the angles $\theta,\phi$. 

The radial functions $P_{n,\kappa}(r)$ (large component) and $Q_{n,\kappa}(r)$ (small component) are obtained by solving numerically the radial Dirac equations, given by
\begin{eqnarray}
\label{radials}
\nonumber
\left[\varepsilon_{n,\kappa}+\left(u_\mathrm{cap}(r)+\frac{1}{r}\right)\right]P_{n,\kappa}(r)+c\left(\dderr-\frac{\kappa}{r}\right)Q_{n,\kappa}(r)&=&0,\\
\left[\varepsilon_{n,\kappa}+2c^2 -\left(u_\mathrm{cap}(r)-\frac{1}{r}\right)\right]Q_{n,\kappa}(r)-c\left(\dderr+\frac{\kappa}{r}\right)P_{n,\kappa}(r)&=&0,
\end{eqnarray}
where  $\varepsilon_{n,\kappa}=E-c^2$. Typically, Eq.~(\ref{radials}) is solved using a $L^2$-basis approximation where Dirichlet boundary conditions are imposed \cite{Grant06}. For this reason, in order to prevent unphysical reflections during propagation, a complex absorbing potential (CAP) can be incorporated \cite{Riss93}. In this work, the CAP is added to the Dirac equation as a \emph{scalar potential} \cite{Greiner73} following  the implementation made by Ackad and Horbatsch in Ref. \cite{Ackad07,Horbatsch07,Horbatsch07b}. Thus, the CAP is defined as
\begin{equation}
    u_\mathrm{cap}(r)=\left\{\begin{array}{rcr}
        0&\mathrm{if} & r\leq r_\mathrm{cap},   \\
        -i\eta(r-r_\mathrm{cap})^2&\mathrm{if} & r>r_\mathrm{cap},
    \end{array}\right.
\end{equation}
being $\eta$ a positive parameter that determines the strength of potential. Due to the presence of the CAP the Dirac Hamiltonian becomes non-Hermitian and continuum states present an imaginary part. As a consequence, the Hermitian inner product is not satisfied in this basis. Nevertheless, the resulting complex symmetric Hamiltonian enables a redefinition of the inner product \cite{Greenman10}. In practice, one doesn't take the complex conjugate of the radial coordinate in the left vector when computing matrix elements. Then, the transition matrix elements in Eq.~(\ref{FKDL}) and in Eq.~(\ref{FKDV}) will be complex in general.  On the contrary, the calculation of physical expectation values should be real. Therefore, matrix elements in Eq.~(\ref{ZMKDL}) and in Eq.~(\ref{VMKDV}) have been evaluated in an \emph{inner region}, which is not affected by the CAP:  $0<r_\mathrm{inner}<r_\mathrm{cap}$, and where the Hermitian conjugate on the radial coordinate has been used on left vector. Thus, for two given states $\Phi_{K'}({\bf r})$ and $\Phi_{K}({\bf r})$, the transition matrix elements $\tilde{M}_{K',K}^{DL}(t)$ and $\tilde{M}_{K',K}^{DV}(t)$, evaluated in the inner region, are given by
\begin{eqnarray}
\label{Element_DL}
\nonumber
\tilde{M}_{K',K}^{DL}(t)&=&E(t)\langle\Phi_{K'}|z|\Phi_{K}\rangle\\
\nonumber
&=&E(t)\langle X_{\kappa',j',m'}|C^1_0| X_{\kappa',j',m'}\rangle\\
& &\times\int_0^{r_\mathrm{inner}}r[P_{n',\kappa'}^*(r)P_{n,\kappa}(r)+Q_{n',\kappa'}^*(r)Q_{n,\kappa}(r)]\der r,
\end{eqnarray}
where $C_{0}^{1}=\sqrt{4\pi/3}\;Y_{1}^{0}(\Omega)$ is a renormalized spherical harmonic in Racah's notation and the spin-angular coefficient is defined as 
\begin{eqnarray}
\nonumber
\langle X_{\kappa',j',m'}|C_0^1| X_{\kappa',j',m'}\rangle&=&(-1)^{m-\frac{1}{2}}[(2j+1)(2j'+1)]^{1/2}\\
& &\times\left(\begin{array}{ccc}
    j & 1 &j'\\
    -\frac{1}{2} &  0&\frac{1}{2}  
 \end{array}\right)\left(\begin{array}{ccc}
    j & 1 &j'\\
    -m&  0&m  
 \end{array}\right),
\end{eqnarray}
and 
\begin{eqnarray}
\label{Element_DV}
\nonumber
\tilde{M}_{K',K}^{DV}(t)&=&cA(t)\langle\Phi_{K'}|\alpha_z|\Phi_{K}\rangle\\
\nonumber
                &=&-icA(t)\left[\int_0^{r_\mathrm{inner}}Q_{n',\kappa'}^*(r)P_{n,\kappa}(r)\der r \langle X_{-\kappa',j',m}|\sigma_z| X_{\kappa,j,m}\rangle \right.\\
                & &\left.-\int_0^{r_\mathrm{inner}}P_{n',\kappa'}^*(r)Q_{n,\kappa}(r)\der r \langle X_{\kappa',j',m}|\sigma_z| X_{-\kappa,j,m}\rangle\right],
\end{eqnarray}
where the spin-angular coefficient is given by the general formula 
\begin{eqnarray}
\nonumber
\langle X_{\kappa',j',m}|\sigma_z| X_{\kappa,j,m}\rangle&=&(-1)^{j'-j+\ell_\kappa-m+\frac{1}{2}}[(2j'+1)(2j+1)]^{1/2}(s||\boldsymbol{\sigma}||s)|_{s=\frac{1}{2}}\\
\nonumber
& &\times \delta_{\ell_{\kappa'},\ell_\kappa}\left\{\begin{array}{ccc}
   \frac{1}{2}  & j' &\ell_\kappa  \\
    j &\frac{1}{2}  &1
     \end{array}\right\}\left(\begin{array}{ccc}
    j&1 &j'  \\
     m&0&-m 
\end{array}\right),
\end{eqnarray}
where the reduced matrix element is $(s||\boldsymbol{\sigma}||s)|_{s=\frac{1}{2}}=\sqrt{6}$ \cite{Grant06}. Finally, the transition matrix elements ${M}_{K',K}^{DL}(t)$ and ${M}_{K',K}^{DV}(t)$, required in Eq.~(\ref{FKDL}) and in Eq.~(\ref{FKDV}), are computed in a similar way than those given by Eq.~(\ref{Element_DL}) and Eq.~(\ref{Element_DV}), but where the radial integrals run from zero to infinity ($r_\mathrm{max}$ in the confined numerical simulation) and without complex conjugation of the left radial functions.

\subsection{Removing the negative energy states from the propagation}
The solution of the Dirac equation is composed by two sets of solutions: the positive energy states and the negative energy states \cite{Sakurai67}. The possibility of removing the negative energy states from the basis, when positron-electron pair creation is energetically out of reach, is very tempting from a computational point of view. The basis is reduced by a factor of two and less demanding propagation in time is necessary. However, only together, the positive and the negative energy solutions form a complete set. Already in the fifties, Furry~\cite{furry:51} showed that when a spectral method is used to evaluate a time-dependent perturbation beyond the lowest order, both positive and negative energy states are needed to expand the intermediate ``virtual'' states correctly. In addition, if the perturbation is described by a non-diagonal operator with respect to the large and the small components of the wave function, it can be shown that the contribution of the negative energy states is of the same order of magnitude as the contribution corresponding to the positive energy states. This has been discussed in some detail in connection with contributions beyond the dipole approximation by  Selst{\o} et al.\cite{Jakob09}, and, in relation with multi-photon ionization, within the dipole approximation, by Vanne and Saenz \cite{Vanne12}.  The velocity form of the dipole operator is indeed non-diagonal with respect to the large and the small component and multi-photon contributions will in this gauge not be correctly represented with just the positive energy spectrum. The length form, on the other hand, is diagonal and the contribution of negative energy states is greatly suppressed. In fact, it will be suppressed beyond the leading relativistic contribution with, at least, one order of the fine-structure constant ($\alpha_{fs} \approx 1/137$). This was also demonstrated numerically in Ref. \cite{Vanne12}. In this work, we have thus chosen to propagate the TDDE in length gauge in order to be able to exclude the negative energy states.  

\subsection{Implementation in a B-spline basis set}
In order to solve Eq. (\ref{radials}), the large and the small components are expanded in B-splines \cite{DeBoor01} basis sets  as in Ref.~\cite{Kjellsson17}. As suggested by Froese-Fischer and Zatsarinny in Ref. \cite{Charlotte08}, we use different polynomial orders for the large and small components in order to remove the so-called \emph{spurious states}  known to appear in the numerical spectrum of the Dirac Hamiltonian after discretization, i.e.
\begin{eqnarray}
\nonumber
 P_{n,\kappa}(r)&=&\sum_{i=1}^{n_s'}a_iB_i^{k_s'}(r),\\
 Q_{n,\kappa}(r)&=&\sum_{j=1}^{n_s}b_jB_j^{k_s}(r),
\end{eqnarray}
where the dimensions of the basis are defined by
$n_s'$ and $n_s=n_s'+1$
 and the order of the B-splines by 
$k_s'$ and $k_s=k_s'+1$. 

In this work, both B-spline sets have been defined on the same sequence
of increasing knot points while the boundary knots have been chosen to be either
$k_{s}$- or $k_{s}'$-fold degenerate, e.g. $r_{1}=r_{2}=...=r_{k_{s}}=r_{\mathrm{min}}$ and $r_{n_{s}+1}=r_{n_{s}+2}=...r_{n_{s}+k_{s}}=r_{\mathrm{max}}$. In order to ensure the zero boundary conditions of $P_{n,\kappa}(r)$ and $Q_{n,\kappa}(r)$ at $r=r_{\mathrm{min}}$ and $r=r_{\mathrm{max}}$, the first and the last B-splines, in both sets, were removed from the calculation. Converged results were obtained using the following parameters: $k_s=8$
with a spatial grid of $n_p=247$ unique knot points values, where the last $50$ knots describe the outer region  ($r>r_\mathrm{cap}$) where the CAP is non-zero, and with $r_\mathrm{inner}=192$ au. This grid generates a total number of $n_s=251$ B-splines for the small component and $250$ for the large component that vanish at the start and the end of the grid. Given this set of B-spline parameters, one obtains 250 positive energy states and 251 negative energy states per spin-angular symmetry.

\section{Numerical validation of the theory: ATAS on Hydrogen atom \label{validation_results}}
\begin{figure}[hb]
\centering
\includegraphics[width=0.8\textwidth]{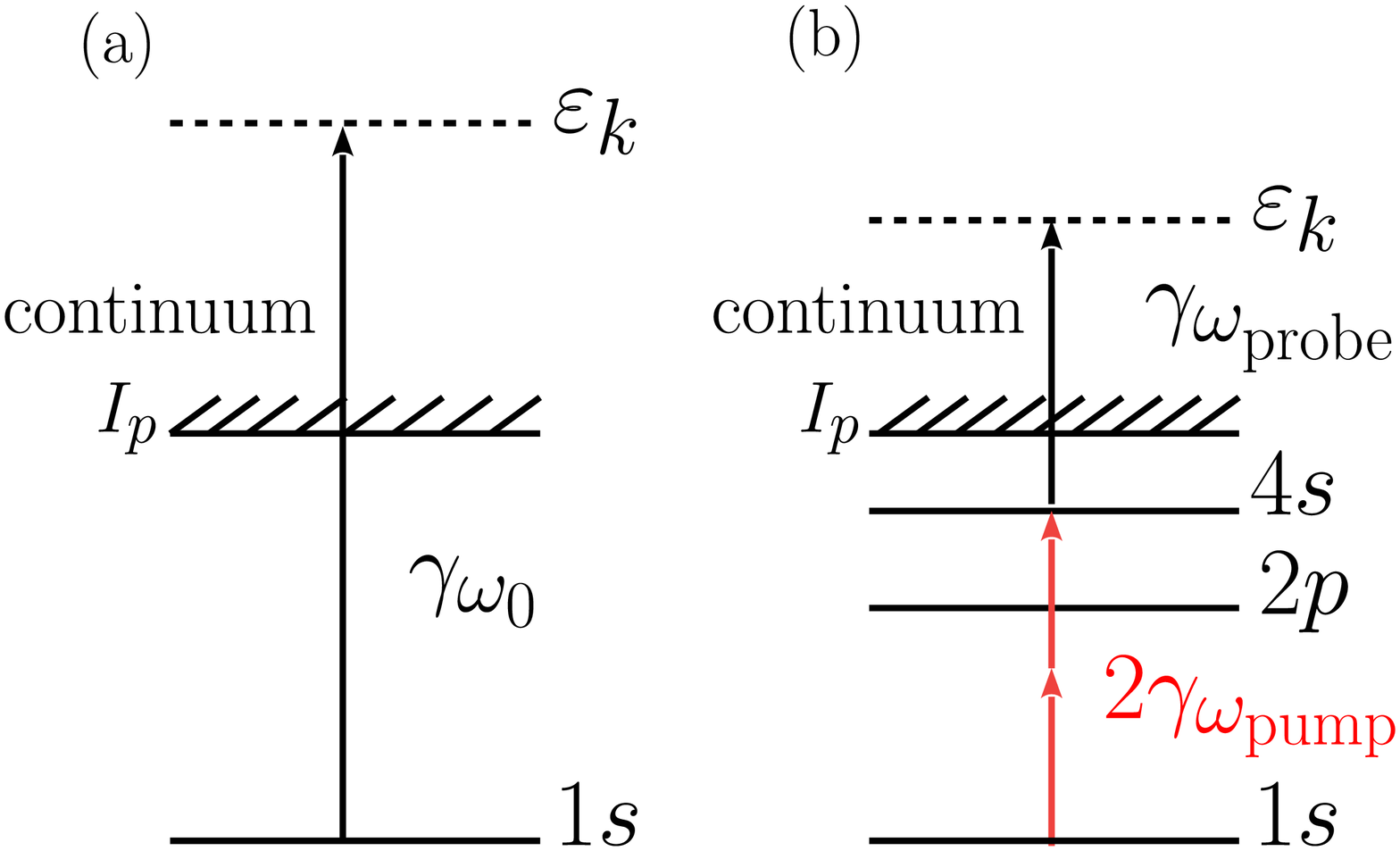}
\caption{ATAS scenarios in Hydrogen atom explored in this paper. Panel (a) shows the one-photon ionization scenario where the reaction is given by $H(1s)+\gamma_{\omega_0}\to H^++ e^-$. In this case, the carrier frequency $\omega_0$ is bigger than the ionization potential $I_p$. In panel (b) the multi-photon ionization process through a resonant state is given. The total reaction can be split into two sub-reactions by the addition of a sufficiently long time delay $t_0$ between the pump and the probe pulses. First, due to the absorption of two pump photons $\gamma_{\omega_\mathrm{pump}}$, the atom is excited from the $1s$ to the $4s$ state. Subsequently, the atom is ionized from the $4s$ state by absorption of one probe photon as follows, $H(4s)+\gamma_{\omega_\mathrm{probe}}\to H^++ e^-$. In this complex scenario, we have imposed: $\omega_\mathrm{pump}<\omega_\mathrm{probe}<I_p$.\label{scenarios}}
\end{figure}

In order to show the validity of the relativistic transient absorption equations derived previously in this work, we have decided to check the non-relativistic limit of the theory. In the non-relativistic limit, results should be very similar to those obtained by the non-relativistic transient absorption theory. For this reason, two different ATAS scenarios in Hydrogen atom have been simulated and compared with non-relativistic calculations. Fig.~(\ref{scenarios}) shows the two ATAS scenarios investigated in this work.

TDSE calculations have been performed following the same spectral method implemented for the TDDE propagation. In addition, the non-relativistic states were expanded onto a B-spline basis set and a CAP was introduced. The B-splines parameters were chosen to be similar to the B-spline basis set defined for the relativistic large component. We included all spin orbitals with angular momenta up to $\ell_\mathrm{max}=10$ and, in order to speed the propagation, the high energy components of the spectrum were not taken into account without compromising results, $\varepsilon_\mathrm{max}=30$ au. The adopted CAP strength was $\eta=5\times 10^{-4}$ and $\Delta t=5\times10^{-3}$ au. 
 
In this work, laser fields are chosen to be Gaussian shaped pulses. The vector potential is defined as 
\begin{equation}
\label{ap}
A(t)=A_0\cos[\omega(t-t_0)]\exp\{-a(t-t_0)^2\},
\end{equation}
where $A_0$ is the envelope amplitude, $\omega_0$ is the carrier frequency and $a=2\ln(2)/\tau_e^2$ is inversely proportional to the squared envelope duration, $\tau_e$, expressed in full-width at half-maximum (FWHM) note that the vector potential approaches zero in both positive and negative infinity in time. The associated electric field is given by
\begin{eqnarray}
\label{er}
\nonumber
E(t)=-&A_0\;\omega&\cos[\omega(t-t_0)]\exp\{-a(t-t_0)^2\} \\
     +&A_0\;b\;(t-t_0)&\sin[\omega(t-t_0)]\exp\{-a(t-t_0)^2\},
\end{eqnarray}
where $b=2a$ and $t_0$ is the time delay.

\subsection{One-photon ionization}
The first ATAS scenario investigated in this work is the one-photon ionization of Hydrogen, \emph{see} panel (a) in Fig.(\ref{scenarios}). The following set of laser parameters has been used: $ A_0 = 2\times10^{-3}$ au, $\omega_0=1.5$ au and $\tau_e=10$ au.

\begin{figure}[ht]
\begin{center}
\includegraphics[width=0.7\textwidth]{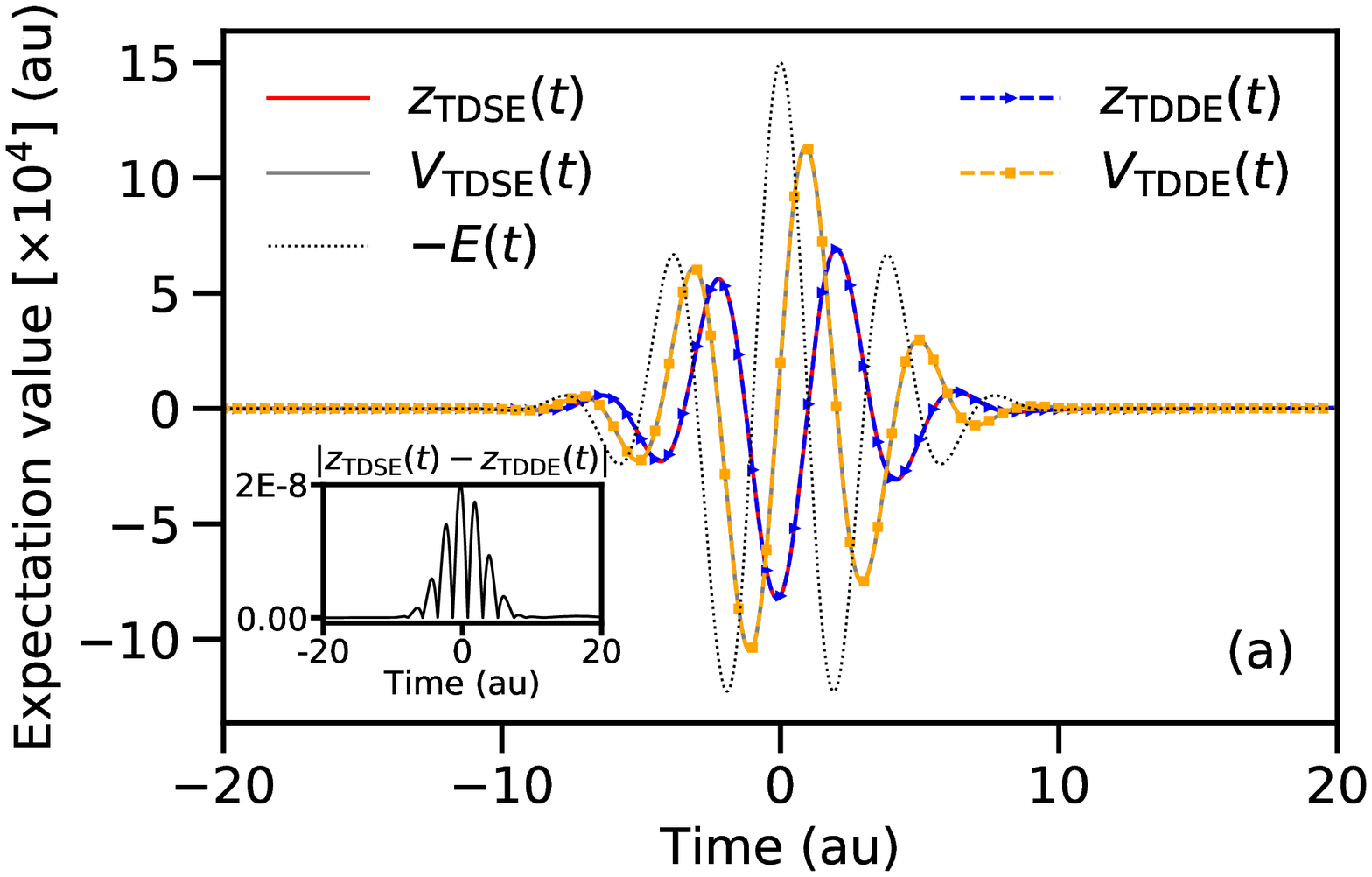}
\includegraphics[width=0.7\textwidth]{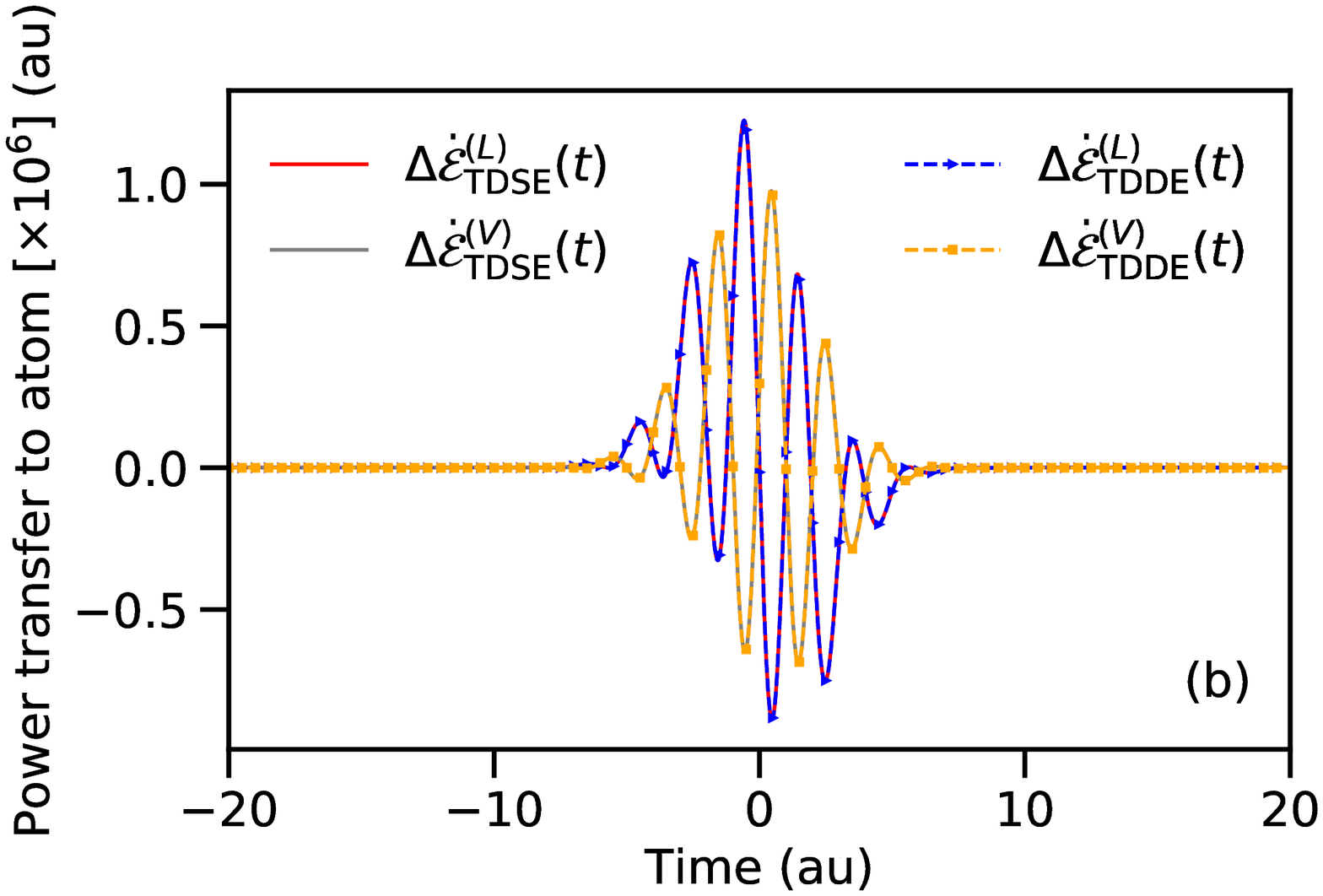}
\includegraphics[width=0.7\textwidth]{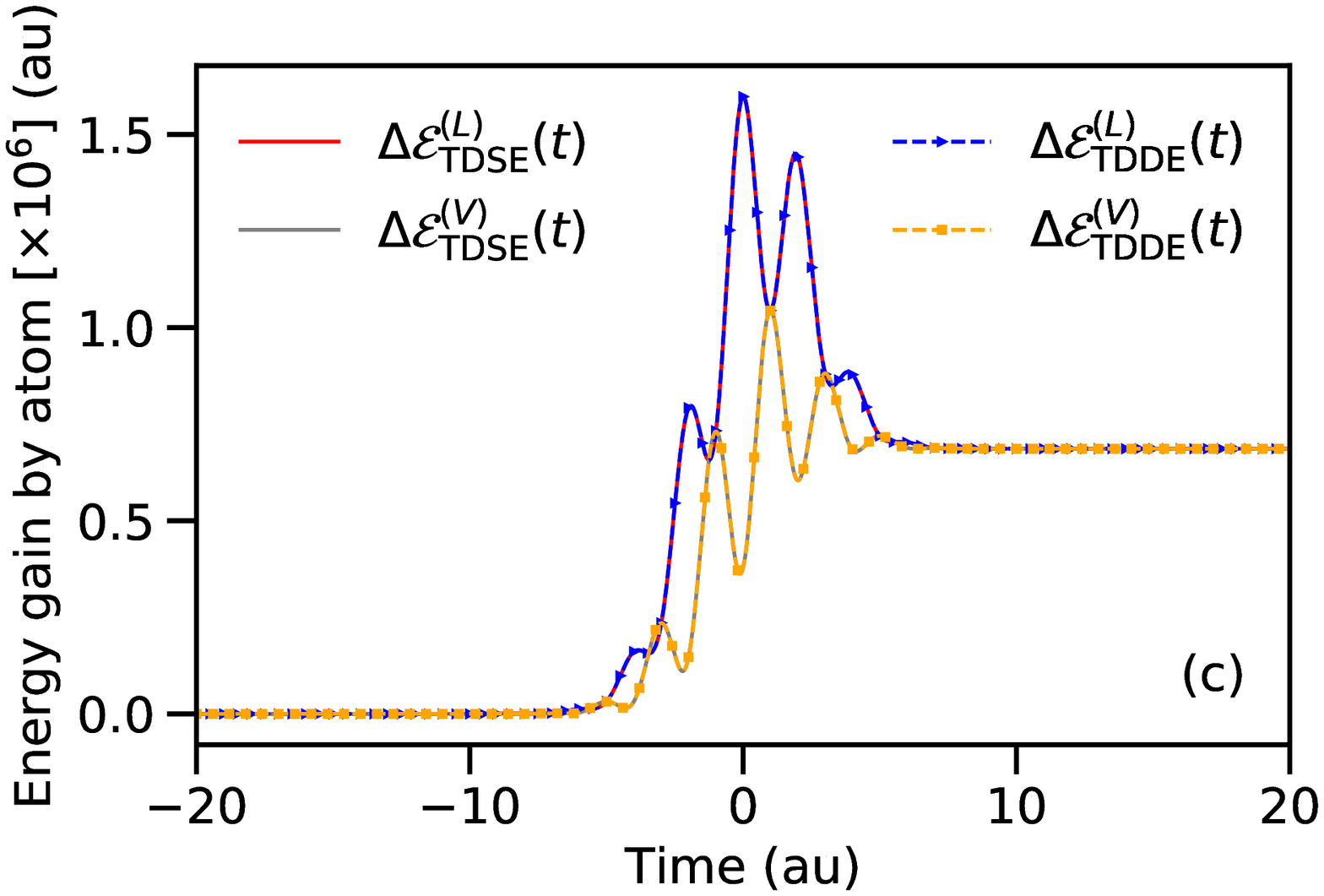}
\caption{In panel (a) time-dependent expectation values of electron position and velocity in Hydrogen atom, induced by the laser pulse, are shown. Panel (b) shows the time-dependent power delivered to the atom computed in length and velocity forms. In panel (c), the integrated power (energy gain) as a function of time is shown.\label{one-photon_time-dependet}}
\end{center}
\end{figure}

In panel (a) of Fig.(\ref{one-photon_time-dependet}), time-dependent expectation values of position and velocity of the electron are compared. The inset Figure shows the difference between the electron position computed with TDSE and TDDE. This difference is found to be in the domain of the numerical error introduced by the propagator for $\Delta t=5\times10^{-3}$ au \cite{Leforestier91}. Thus, differences between TDSE and TDDE calculations can not be resolved. Moreover, one can observe that expectation values oscillate only when the interaction with the laser field is non-zero in this photoionization scenario. The electron position presents a phase lag of $\sim\pi$ with respect to the driven force, $-E(t)$. This $\pi$-shift implies a strongly over-driven system. Furthermore, as expected for an oscillating system, the velocity is shifted by $\sim\pi/2$ in between the force and the position.

In panel (b) of Fig.(\ref{one-photon_time-dependet}), the instantaneous power delivered to the atom is presented in its length and velocity forms. The quantities showed here correspond to non-relativistic equations, Eq.(\ref{power_SL}) and Eq.(\ref{power_SV}), and to the relativistic equations, Eq.(\ref{power_DL}) and Eq.(\ref{power_DV}). As we are placed at the non-relativistic limit of the TDDE, results overlap. Likewise, as Dahlstr\"om \emph{et al} showed in Ref. \cite{Marcus17}, we notice that the time-dependent instantaneous power is gauge dependent\footnote{We have found that the labels for ``Velocity'' and ``No $A^2(t)/2$'' curves are reversed in Figure 1 in Ref. \cite{Marcus17}}. Finally, in panel (c), the integrated power (energy gain by the atom) is shown as a function of time. Here again, during the action of the laser field, this quantity is gauge dependent. However, after the laser pulse is over, the total integrated power is identical in both gauges.

\begin{figure}[ht]
\begin{center}
\includegraphics[width=0.7\textwidth]{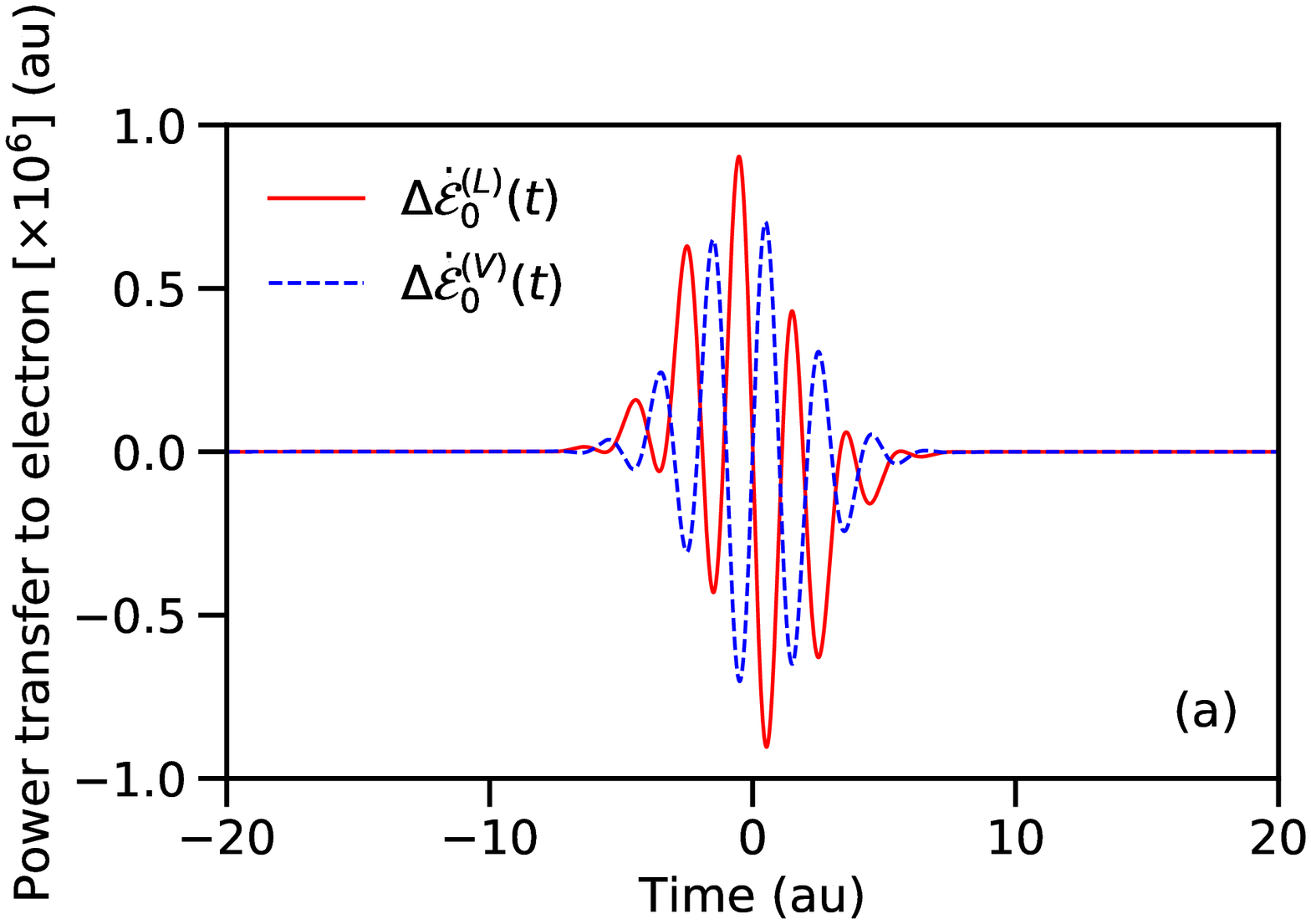}
\includegraphics[width=0.7\textwidth]{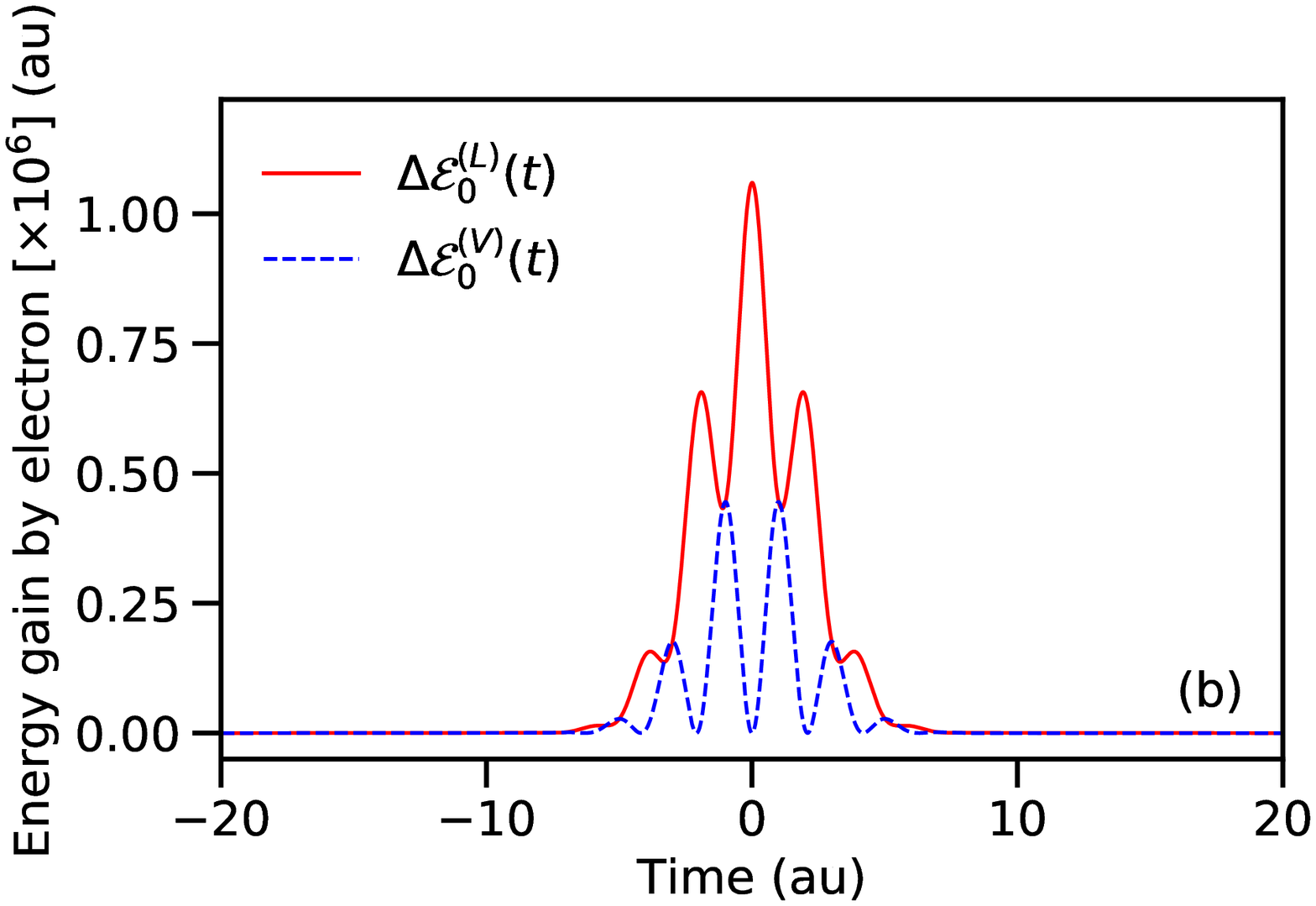}
\caption{In panel (a), the time-dependent instantaneous power transfer to a non-relativistic free electron is presented. In panel (b), the integrated power (energy gain) as a function of time is shown. \label{free_electron}}
\end{center}
\end{figure}

In order to understand the behavior of these time-dependent quantities, we have decided to eliminate the contribution coming from the free electron motion. In Fig.~(\ref{free_electron}), the instantaneous power and the integrated power, as a function of time, delivered to a non-relativistic free electron (used here as a reference physical system) are shown in length and in velocity gauge. In panel (a), strong oscillations in the power transfer to the free electron are observed. In panel (b), the energy gain by the free electron also oscillates, but when the laser pulse is over the total energy absorbed by the electron is zero ($\Delta\mathcal{E}_0^{(L)}=\Delta\mathcal{E}_0^{(V)}=0$) in both gauges as result of Eq.(\ref{ap}).  

\begin{figure}[ht]
\begin{center}
\includegraphics[width=0.7\textwidth]{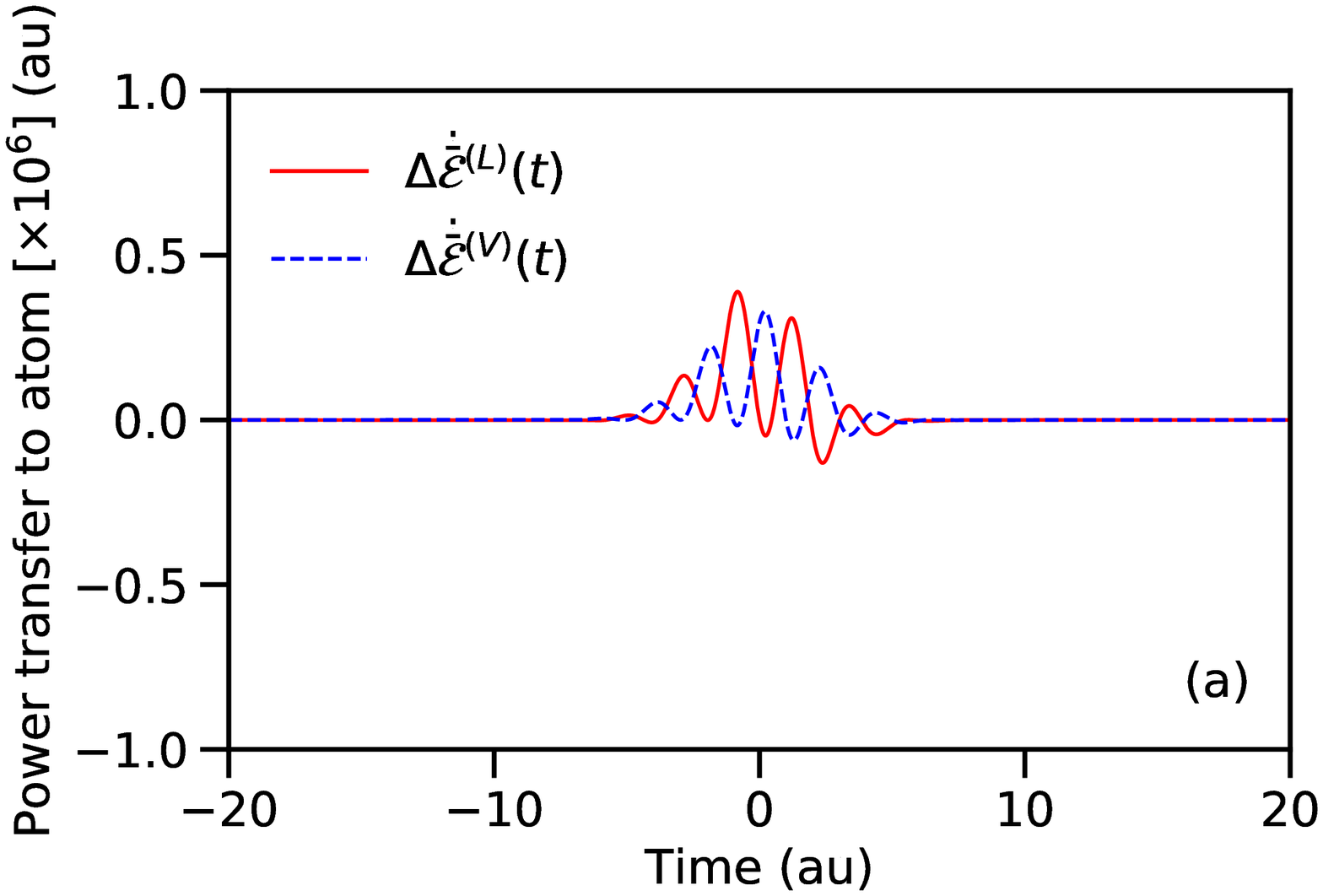}
\includegraphics[width=0.7\textwidth]{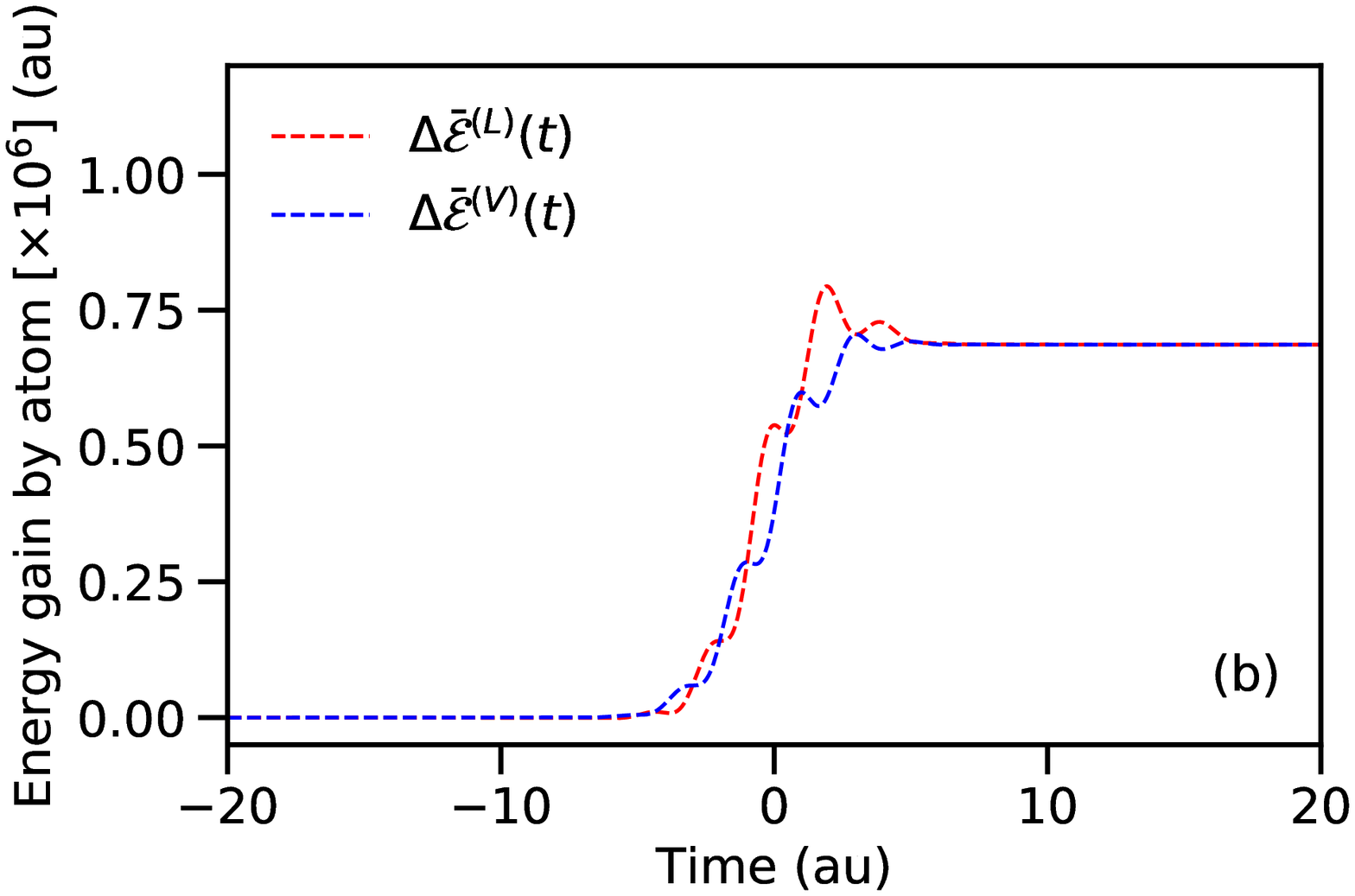}
\caption{In panel (a), the relative time-dependent instantaneous power is shown. In panel (b), the relative integrated power (relative energy gain) as a function of time is given. \label{relative_free_atom}} 
\end{center}
\end{figure}

In Fig.~(\ref{relative_free_atom}), the relative instantaneous power, $\Delta\dot{\bar{\mathcal{E}}}(t)$, and the relative integrated power, $\Delta\bar{{\mathcal{E}}}(t)$ are given as a function of time. Numerical identical results were obtained in the relativistic calculation (not shown here). As we can see in panel (a), the time-dependent oscillations are less significant after removing the free electron contribution and the power transfer to the atom is mostly positive during the interaction with the pulse. Thus, in panel (b), the energy gained by the atom increases gradually and smoothly in comparison with panel (c) in Fig.~(\ref{one-photon_time-dependet}). Removing the free electron contribution allows us to have a better physical view of the absorption process. However,  quantities are not gauge invariant in time under the action of the laser pulse. An invariant formulation of the theory may be possible and can be explored in the future using the definition of the time-dependent energy operator as a starting point, \emph{see} for example Ref.~\cite{Kobe78,Kobe82} and references therein. 

\begin{figure}[h]
\begin{center}
\includegraphics[width=0.7\textwidth]{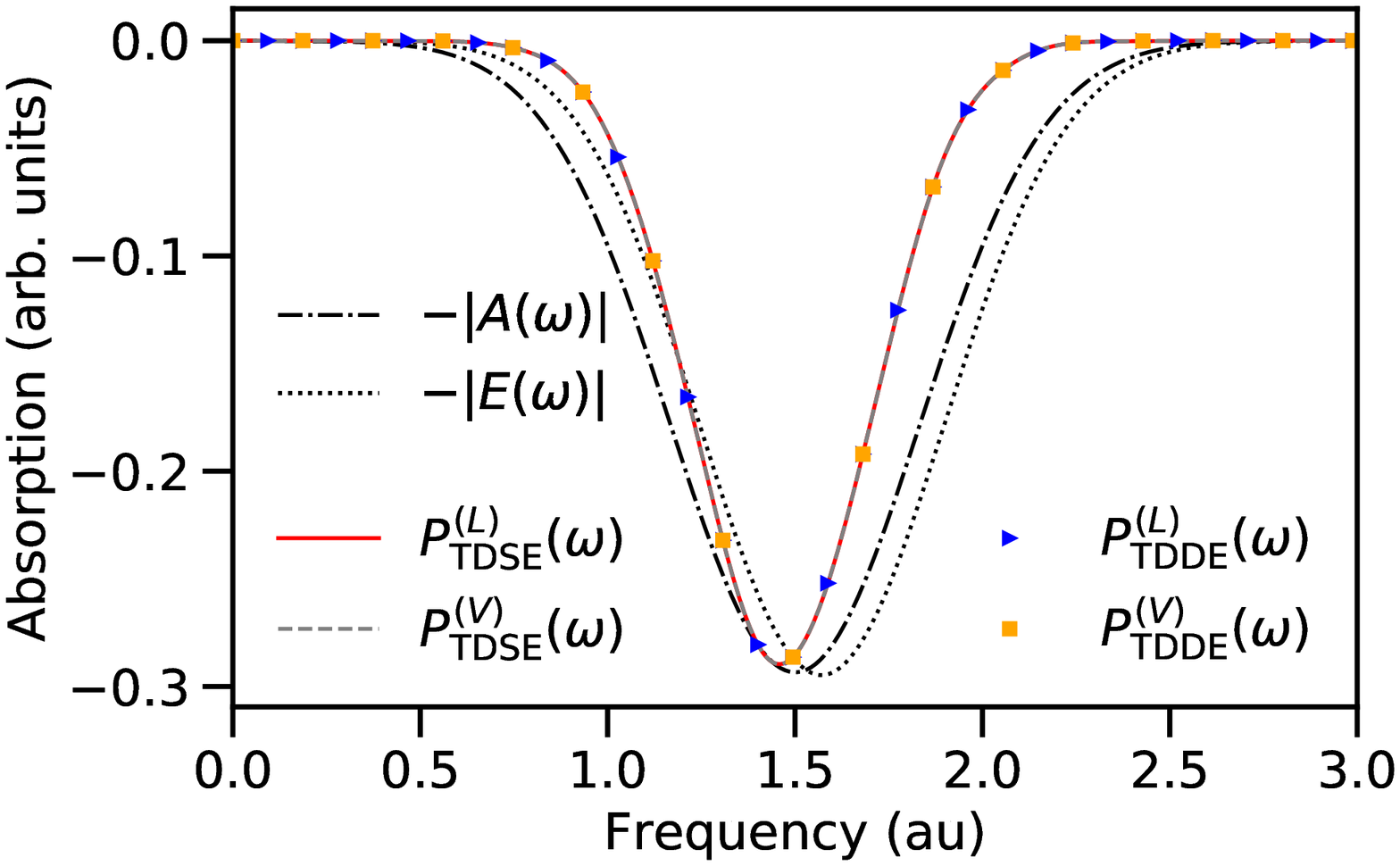}
\caption{Energy-resolved power delivered to the Hydrogen atom. The data has been interpolated between discrete values by a cubic spline function.\label{one-photon_absorption}}
\end{center}
\end{figure}

Finally, the one-photon energy-resolved power is shown in Fig.(\ref{one-photon_absorption}). The  normalized Fourier transforms of the electric field and the vector potential are also given for comparison. The absorption peak is located close to the central photon energy, $\omega_0=1.5$ au, in agreement with the vector potential, while the electric field is displaced to a higher energy. This fact is due to the short duration of the pulse we used here. The energy-resolved power is gauge invariant in both relativistic and non-relativistic theories.

\subsection{Multi-photon ionization}
Typically, in ATAS experiments, several laser fields are involved. Then, by varying the time delay $t_0$ between the pulses, coherent electron excitations may be controlled.  In this second scenario, we have decided to investigate the multi-photon ionization of the Hydrogen atom induced by a pump pulse and a probe pulse, \emph{see} panel (b) in Fig.(\ref{scenarios}). The chosen laser parameters were; for the pump pulse: $A_0=5\times10^{-3}$ au, $\omega_\mathrm{pump}=0.234245$ au, $\tau_e=1000$ au, for the probe pulse: $A_0=1\times10^{-3}$ au, $\omega_\mathrm{probe}=0.3$ au, $\tau_e=100$ au. The time delay of the probe with respect to the pump was $t_0=4000$ au.

\begin{figure}[h]
\begin{center}
\includegraphics[width=.7\textwidth]{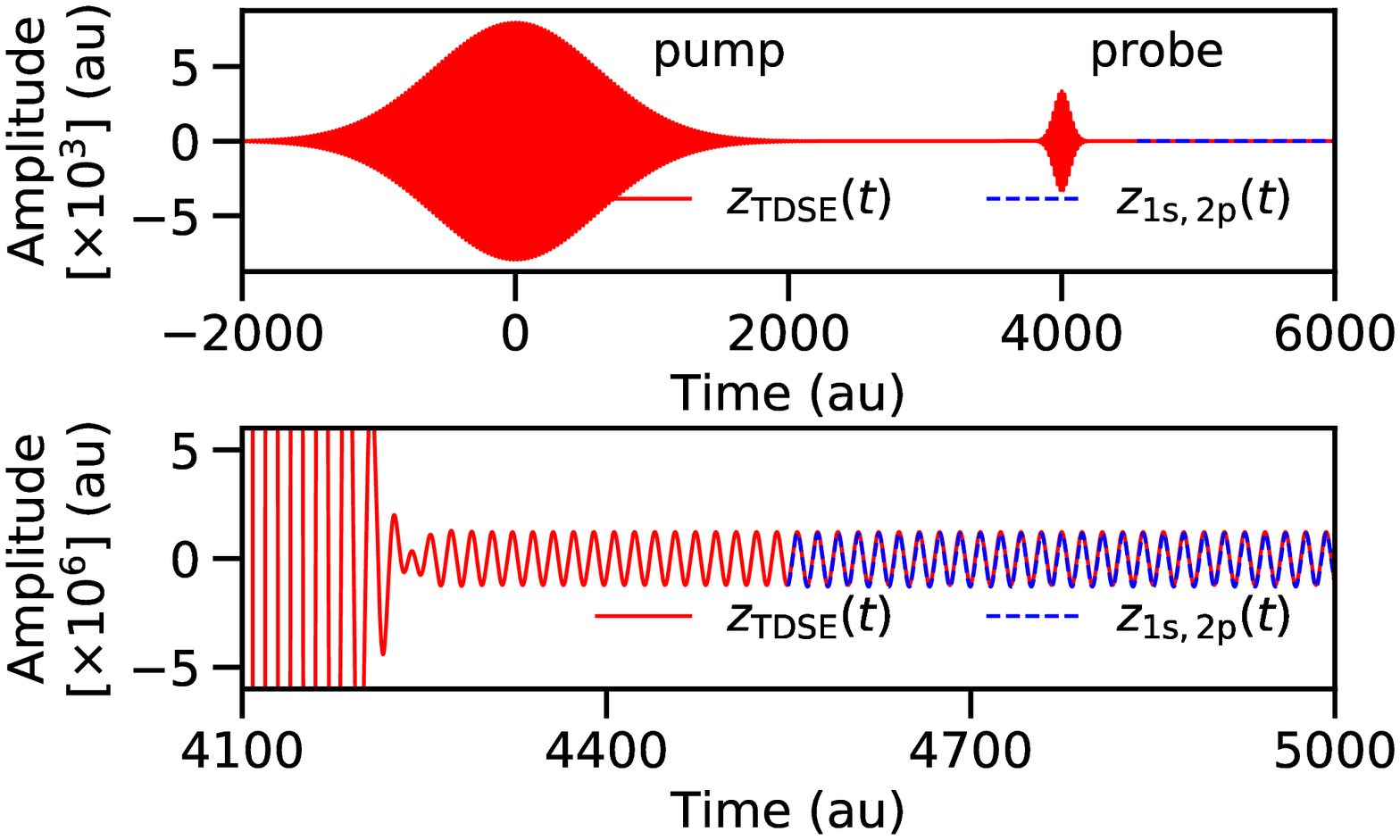}
\caption{The red solid line represents the time-dependent expectation value of the electron position induced by the pump and the probe laser fields. The blue dashed line represents the field-free evolution of the $1s$--$2p$ spectral component of the electron position.\label{multi-photon_dipole}}
\end{center}
\end{figure}

The former parameters have been chosen carefully to ensure the scenario described in the panel (b) of Fig.(\ref{scenarios}), where two pump photons $\gamma_\mathrm{pump}$ excite the system from the ground state $1s$ to the excited state $4s$, and then, after waiting a time equals to $t_0$, the atom is ionized from the $4s$ state by absorption of one probe photon, $\gamma_\mathrm{probe}$. In Fig.(\ref{multi-photon_dipole}) the time-dependent expectation value of the electron position is shown together with the field-free propagation of the $1s$-$2p$ spectral component. In the upper panel, the time-dependent expectation value is shown. The response to the pump and to the probe pulses can be easily appreciated. However, a small amount of electron density is moved from the ground state to the excited state $2p$ due to the spectral width of the probe pulse. Indeed, this coherence can be detected by analyzing the time-evolution of the spectral components of the time-dependent electron position. By comparing with the field-free propagation of the $1s$-$2p$ spectral component, we can see that it corresponds to an excitation from the ground state to the $2p$ state by absorbing a probe photon $\gamma_\mathrm{probe}$. In order to analyze the information encoded in the total response, these kind of unceasing oscillations may be removed as their contributions to the energy-resolved spectrum are not of interest. Different filter methods are proposed in the literature, \emph{see} for example Re. \cite{Wu16} and references therein. However, we have decided to follow the procedure given by Petersson \emph{et al} in Ref. \cite{Petersson17}, where a time-dependent smooth step function is introduced to split the time-dependent expectation value of the position into a short range and a long range domain. In Ref. \cite{Petersson17}, it is also shown that the long range part of the expectation value, corresponding to perpetual oscillations, can be accurately expressed in terms of the field-free spectral components.

\begin{figure}[h]
\begin{center}
\includegraphics[width=.7\textwidth]{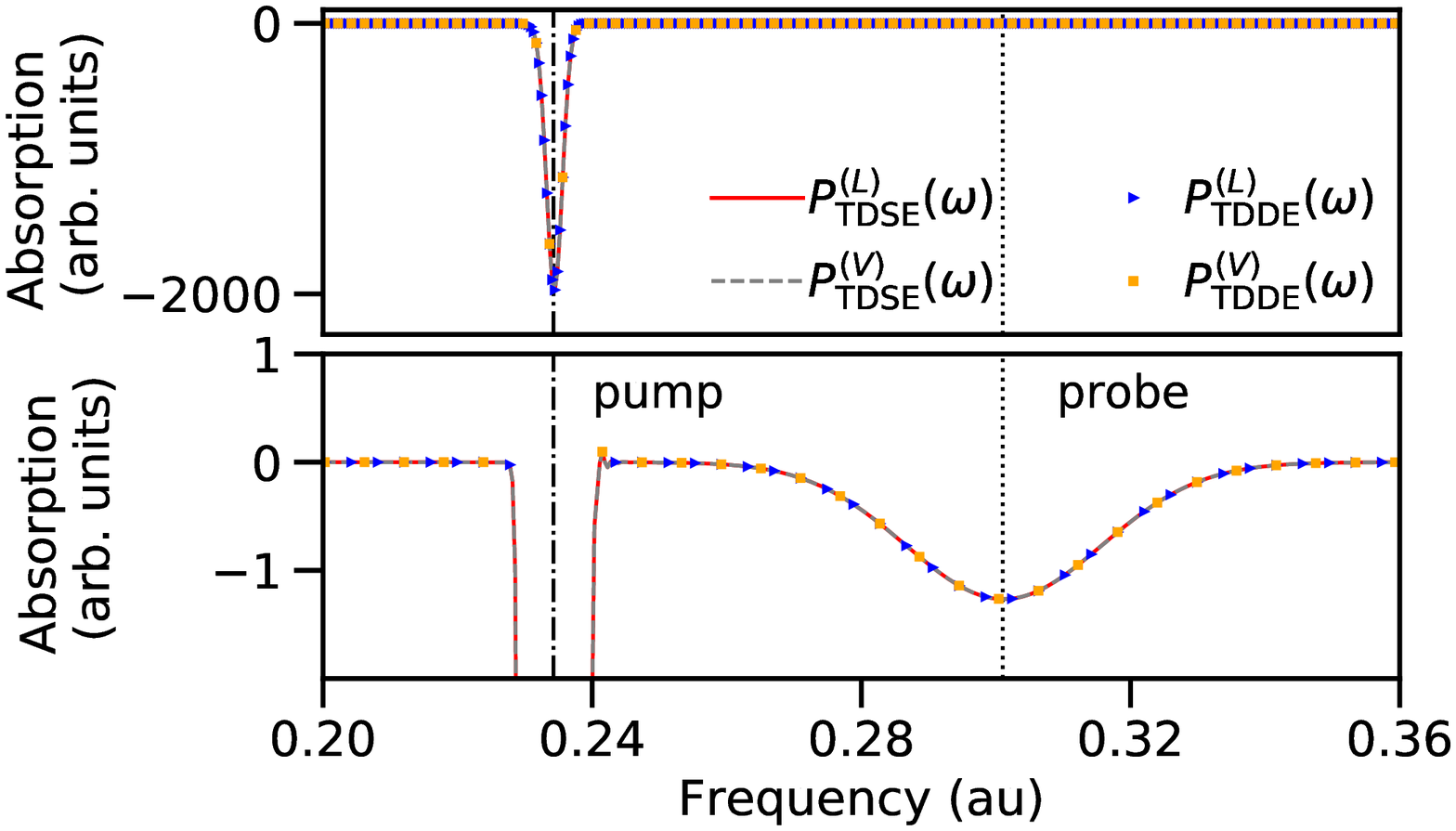}
\caption{Energy-resolved power delivered to the Hydrogen atom by the pump and the probe pulses in the multiphoton scenario.\label{multi-photon_power}}
\end{center}
\end{figure}

As a step function, we have used the inverse of the cumulative distribution function, where 
\begin{equation}
    f(t)=\frac{1}{2}\left[1+\mathrm{erf}\left(\frac{t-\mu}{\sigma\sqrt{2}}\right)\right],
\end{equation}
with $\mu=5800$ au and $\sigma=200$ au. After applying this filter to the time-dependent expectation value of the position, given in Fig.~(\ref{multi-photon_dipole}), the energy-resolved power delivered to the Hydrogen atom without the long range contribution is given in Fig.(\ref{multi-photon_power}). Non-relativistic and relativistic calculations overlap, verifying the validity of Eq.(\ref{ins_rl}) and Eq.(\ref{ins_rv}) and gauge invariant of the multi-photon process.

\section{Conclusion\label{conclusion}}
In this article we have presented a relativistic transient absorption theory that can be seen as an extension of the well-established non-relativistic theory. 
The relativistic energy-resolved power was derived in both length and velocity gauges. Our derivations were validated by numerical simulation of the Hydrogen atom in two different pump--probe scenarios. The simulations confirm that the energy-resolved absorption is a gauge invariant quantity. The derivations presented here can be used for the development of a more general ATAS method for studying heavy atoms. Since our present theory is based on one-particle relativistic expectation values, we suggest the possibility to explore relativistic processes, such as the \emph{Zitterbewegung} effect, in the time domain using transient absorption techniques. Finally, in order to make a more physical interpretation of the time-dependent power, which is not a gauge invariant quantity, we have introduced a relative instantaneous power to a free electron. An interesting further development of the present time-dependent theory could consist in a gauge invariant reformulation.

\printbibliography

\end{document}